%% file: stirap.tex
\documentclass[a4paper, pra, amsmath, showpacs, preprintnumbers,
superscriptaddress, twocolumn, sort&compress, floatfix]{revtex4}
\usepackage{times,mathptm}
\usepackage{graphicx, color}

\usepackage{textcomp}
\usepackage[latin1]{inputenc}
\newcommand{\summary}[1]{\mbox{}\marginpar{\raggedright\hspace{0pt}\scriptsize\bfseries[#1]}}
\renewcommand{\summary}[1]{}

\begin{document}

\title{Coherent optical transfer of Feshbach molecules to a lower vibrational state}

\author{K. Winkler}
\affiliation{Institut f\"ur Experimentalphysik, Forschungszentrum
f\"ur Quantenphysik, Universit\"at
  Innsbruck, % Technikerstra{\ss}e 25,
  6020 Innsbruck, Austria}
\author{F. Lang}
\affiliation{Institut f\"ur Experimentalphysik, Forschungszentrum
f\"ur Quantenphysik, Universit\"at
  Innsbruck, % Technikerstra{\ss}e 25,
  6020 Innsbruck, Austria}
\author{G. Thalhammer}
\affiliation{Institut f\"ur Experimentalphysik, Forschungszentrum
f\"ur Quantenphysik, Universit\"at
  Innsbruck, % Technikerstra{\ss}e 25,
  6020 Innsbruck, Austria}

\author{P. v. d. Straten}
\affiliation{Debye Institute, Universiteit Utrecht,
  3508 TA Utrecht, Netherlands}

\author{R. Grimm}
\affiliation{Institut f\"ur Experimentalphysik, Forschungszentrum
f\"ur Quantenphysik, Universit\"at
  Innsbruck, % Technikerstra{\ss}e 25,
  6020 Innsbruck, Austria}
\affiliation{Institut f\"ur Quantenoptik und Quanteninformation,
\"Osterreichische Akademie der Wissenschaften, 6020 Innsbruck,
Austria}

\author{J. Hecker Denschlag}
\affiliation{Institut f\"ur Experimentalphysik, Forschungszentrum
f\"ur Quantenphysik, Universit\"at
  Innsbruck, % Technikerstra{\ss}e 25,
  6020 Innsbruck, Austria}

\date{\today}

\pacs{34.50.Rk, 32.80.Pj, 03.75.Nt, 42.50.Gy}

\begin{abstract}
Using the technique of stimulated Raman adiabatic passage (STIRAP) we have coherently
transferred ultracold $^{87}$Rb$_2$ Feshbach molecules into a more deeply bound vibrational
quantum level. Our measurements indicate a high transfer efficiency of up to 87\%. As the
molecules are held in an optical lattice with not more than a single molecule per lattice
site, inelastic collisions between the molecules are suppressed and we observe long molecular
lifetimes of about 1\,s. Using STIRAP we have created quantum superpositions of the two
molecular states and tested their coherence interferometrically. These results represent an
important step towards Bose-Einstein condensation (BEC) of molecules in the vibrational ground
state.
\end{abstract}

\maketitle

\summary{cold molecules}%

The rapidly growing interest in molecular quantum gases is partially
inspired by the success in the field of ultracold
atoms~\cite{review-nature}. Since molecules have more internal
degrees of freedom than atoms, ultracold molecules lend themselves
to an even larger number of interesting studies in, for instance,
few body collision physics~\cite{Chi05,Sta06,Zah06}, chemistry in
the ultracold regime, high resolution spectroscopy, as well as
quantum computation~\cite{Mil02}. Furthermore,  molecules in their
vibrational ground state are of special interest, because they allow
for the formation of an intrinsically stable molecular BEC.

Current pathways towards ultracold molecules in well-defined quantum states are either based
on sympathetic cooling~\cite{Doy04} or association of ultracold neutral atoms, using
photoassociation~\cite{Jon06} or Feshbach resonances~\cite{Koh06}. The method of Feshbach
ramping has proved especially successful and efficient, but it only produces molecules in the
least bound vibrational level. In order to selectively convert molecules into more deeply
bound states, it has been proposed~\cite{Jak02} to use a sequence of stimulated optical Raman
transitions to step molecules down the vibrational ladder. This process takes place while the
molecules are kept in an optical lattice, which isolates them from each other and thus shields
them from detrimental collisions. Recently, optical transfer of molecules into their
vibrational ground state is demonstrated experimentally using a ``pump-dump'' method without
lattice at a moderate efficiency and selectivity~\cite{Sag05}.
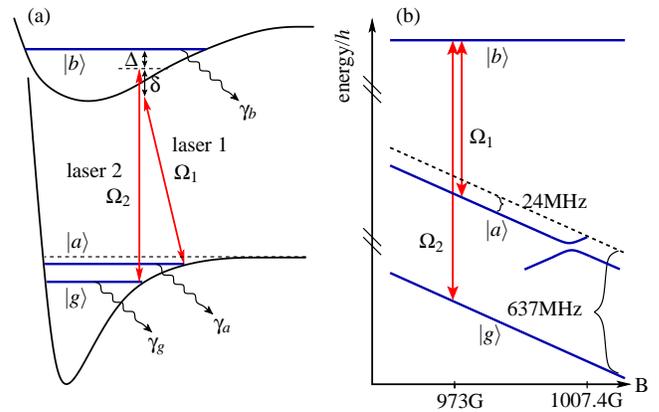
\begin{figure}
\input{figure1}
\caption{(a) Level scheme for STIRAP. Lasers 1, 2 couple the ground state molecular levels
$|a\rangle$, $|g\rangle$ to the excited level $|b\rangle$ with Rabi frequencies $\Omega_{1},
\Omega_{2}$, respectively. $\Delta$ and $\delta $ denote detunings. $\gamma_{a}, \gamma_{b},
\gamma_{c}$ give effective decay rates of the levels. (b) Zeeman diagram of relevant energy
levels. At 1007.4\,G a molecular state crosses the threshold of the unbound two atom continuum
(dashed line) giving rise to a Feshbach resonance. From there this molecular state
adiabatically connects to the last bound vibrational level $|a\rangle$, the state of the
Feshbach molecules.} \label{fig:raman-scheme}
\end{figure}

Here we report the realization of an efficient and highly selective transfer scheme, where an
ensemble of $^{87}$Rb$_2$ Feshbach molecules in an optical lattice is coherently converted to
a deeper bound molecular state via stimulated Raman adiabatic passage (STIRAP). STIRAP is
known as a fast, efficient and robust process for population transfer based on a Raman
transition~\cite{Ber98}. During transfer it keeps the molecules in a dark superposition state,
which decouples from the light and thus suppresses losses due to spontaneous light scattering.
In our proof-of-principle experiment we transfer the Feshbach molecules with a STIRAP pulse
from their last bound vibrational level (binding energy 24\,MHz$\times h$), which we denote
$|a\rangle$, to the second to last bound vibrational level, $|g\rangle$ (see
Fig.~\ref{fig:raman-scheme}a,b). Both levels have a rotational quantum number $l=0$ and a
total spin $F = 2, m_F = 2$. The level $|g\rangle$ is known from previous
experiments~\cite{Wynar,Rom,Tha05,Win05}. It has a binding energy of 637\,MHz$\times h$ at
973\,G and can be conveniently reached via Raman beams generated with an acousto-optic
modulator (AOM). In order to detect the more deeply bound molecules, a second STIRAP pulse
converts the molecules back to the last bound vibrational level, where they are detected as
atoms after dissociation via Feshbach ramping. The complete cycle has an efficiency of 75\%,
indicating a single STIRAP efficiency of 87\%.

%%%%%%%%%%%%%%%%%%%%%%%%%%%%%

\begin{figure}
\includegraphics[width=0.9\columnwidth]{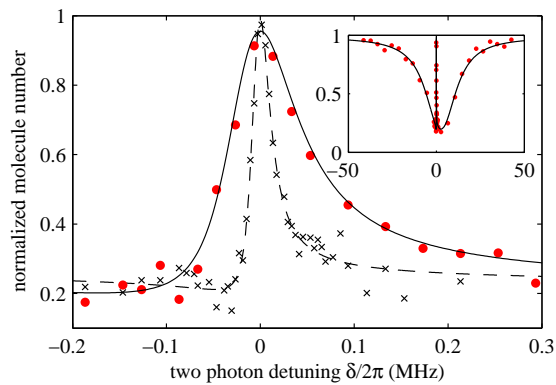}
\caption{Dark resonance. The data show the remaining fraction of Feshbach molecules in state
$|a\rangle$, after subjecting them to a $200\,$\textmu{}s square pulse of Raman laser light in
a narrow range around 0 of the two-photon detuning $\delta$. The inset shows the scan over the
whole line of state $|b\rangle$. The strong suppression of loss at $\delta = 0$ is due to the
appearance of a dark state. The laser intensities are $I_{1} = 2.6\,$mW/cm$^2$, $I_{2} =
13\,$mW/cm$^2$ (crosses), $I_{2}= 51\,$mW/cm$^2$ (filled circles). $\Delta$ is in general
tuned close to zero and for the shown measurements happens to be
$\Delta/2\pi\,\approx\,2.5\,$MHz, which gives rise to the slightly asymmetric line shape of
the dark states. The solid and dashed line are model calculations (see text).}
\label{fig:autler}
\end{figure}

We use essentially the same setup as in  Ref.~\cite{Tha06}. Starting point for the experiments
is a pure ensemble of $2 \times 10^4$ ultracold $^{87}$Rb$_2$ Feshbach molecules which are
held in the lowest Bloch band of a cubic 3D optical lattice. There is no more than a single
molecule per site and the whole molecular ensemble occupies a volume of about $20\times20
\times20\,$\textmu{}m$^3$. The lattice is 50\,$E_r$ deep for molecules
($E_r=2\pi^2\hbar^2/m\lambda^2$, where $m$ is the mass of the atoms and $\lambda= 830.44\,$nm
the wavelength of the lattice laser), suppressing tunneling between sites. The molecular
ensemble is initially produced from an atomic $^{87}$Rb Bose-Einstein condensate (BEC) after
loading it into the lattice, subsequent Feshbach ramping at 1007.40\,G \cite{Vol03} and a
final purification step \cite{Tha06} which removes all chemically unbound atoms. Lowering the
magnetic field to 973\,G transfers the atoms to the adiabatically connected state $|a\rangle$,
which has nearly the same magnetic moment as $|g\rangle$ (see Fig. \ref{fig:raman-scheme}).
This results in an almost magnetic field insensitive Raman transition \footnote{At 973\,G the
dependence of the two-photon resonance on the magnetic field is about 12\,kHz/G, a residual
effect of the level crossing below the Feshbach resonance at 1007.4\,G.}.

In order to efficiently carry out STIRAP, a suitable excited molecular level, $|b\rangle$, has
to be identified (see Fig.~\ref{fig:raman-scheme}). After an extensive search, we finally
chose the electronically excited molecular state $|0_g^-,\nu = 31, J = 0 \rangle$ located
$6.87\,\text{cm}^{-1}$ below the $S_{1/2} + P_{3/2}$ dissociation asymptote~\cite{Fioretti}.
The corresponding line is strong and solitary, i.e. within a 2\,GHz vicinity no other strong
molecular lines are found which could interfere with STIRAP. Coupling to other excited
molecular states leads to loss of the molecules, since these levels typically decay
spontaneously into a variety of undetected vibrational levels in the ground state. Furthermore
it is advantageous that the chosen level $|b\rangle$ has a similar Franck-Condon overlap with
states $|a\rangle$ and $|g\rangle$. It can be shown that this also helps to minimize losses
through off-resonant coupling channels.

With this choice of states $|a\rangle$, $|b\rangle$,
$|g\rangle$, we observe a clear molecular dark resonance when coupling them with resonant
Raman laser light (see Fig. \ref{fig:autler}). The corresponding molecular dark superposition
state shows a long lifetime. This is a necessary precondition for our STIRAP experiments,
because the molecules have to be kept in a dark state during the whole STIRAP process which in
our case typically takes hundreds of \textmu{}s
 \footnote{Note that these STIRAP times are
 several orders of magnitude longer as compared to previous STIRAP
 experiments with thermal molecules~\cite{Ber98}.}.
The Raman laser beams are both derived from a single Ti:Sapphire laser with a short term line
width of less than 1\,MHz. The Ti:Sapphire laser is offset locked relative to the $D_2$-line
of atomic rubidium with the help of a scanning optical cavity, which yields an absolute
frequency stability of better than 5\,MHz. The frequency difference between the two beams is
created with an acousto-optical modulator (AOM) with a frequency of about 307\,MHz in a
double-pass configuration. This allows precise control of the relative frequency difference
between the beams over several tens of MHz and ensures phase-locking. Both beams propagate
collinearly and have a waist of about $290\,$\textmu{}m at the location of the molecular
ensemble. The polarization of the beams is parallel to the direction of the magnetic bias
field of 973\,G.

\begin{figure}
\includegraphics[width=\columnwidth]{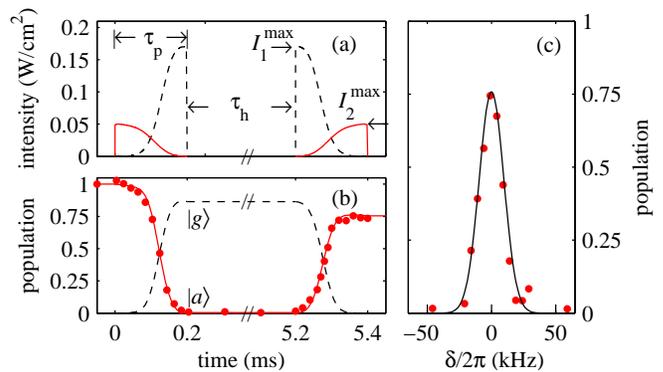}
\caption{STIRAP. (a) Counterintuitive pulse scheme. Shown are laser intensities as a function
of time (laser\,1: dashed line, laser~2: solid line). The first STIRAP pulse with length
$\tau_\text{p} = 200\,$\textmu{}s transfers the molecule from state $|a\rangle$ to state
$|g\rangle$. After a hold time $\tau_\text{h} = 5\,$ms, the second pulse (identical, but
reversed) transfers the molecules back to $|a\rangle$. $I^{\text{max}}_{1,2}$ indicates the
maximal intensity of laser\,1 (2) in the pulse, respectively. (b) Corresponding population in
state $|a\rangle$ (data points, solid line) and state $|g\rangle$ (dashed line).  The data
points are measurements where at a given point in time the STIRAP lasers are abruptly switched
off and the molecule population in state $|a\rangle$ is determined.  For these measurements
$\Delta \approx 0 \approx \delta$. The lines are model calculations (see text). (c) Efficiency
for population transfer from state $|a\rangle$ to state $|g\rangle$ and back via STIRAP as a
function of the two photon detuning $\delta$. The line is a model calculation, showing a
Gaussian line shape with a FWHM width of $\,\approx22\,$kHz.} \label{fig:stirapplot}
\end{figure}

In order to transfer the molecules from state $|a\rangle$ to state $|g\rangle$, we carry out a
STIRAP pulse which consists of a so-called counterintuitive succession of two laser pulses
(see Fig.~\ref{fig:stirapplot}a). We first switch on laser~2 and then ramp its intensity to
zero within the pulse time $\tau_\text{p}$ = 200$\,$\textmu{}s. Simultaneously we ramp up the
intensity of laser~1 from zero to its final value. We fix the ratio of the maximal pulse
intensities of laser$\,1$ and 2 to $I^\text{max}_{2}/I^\text{max}_{1}=1/3.2$ in order to
partially compensate for the unequal Franck-Condon factor of the
 $|a\rangle$--$|b\rangle$ and  $|g\rangle$--$|b\rangle$ transitions.
Ideally, after the first STIRAP pulse all molecules from state $|a\rangle$ should end up in
state $|g\rangle$. In order to determine the population in state $|g\rangle$, we apply, after
a hold time of $\tau_\text{h}\,=\,5\,$ms, a second STIRAP pulse which is the mirror image in
time of pulse\,1. This transfers the molecules back into state $|a\rangle$. We then ramp the
magnetic field over the Feshbach resonance at $1007.4\,$G which dissociates the molecules with
unit efficiency \cite{Tha06} into pairs of atoms. These are subsequently detected with
standard absorption imaging. Fig.~\ref{fig:stirapplot}b shows in a time resolved way how
molecules in state $|a\rangle$ first disappear and then reappear during the course of the
STIRAP sequence. After applying the first STIRAP pulse, no molecules can be observed in state
$|a\rangle$. This is to be expected, since any molecule which is left over in state
$|a\rangle$ at the end of the first STIRAP pulse is in a bright state and will be quickly
removed by resonantly scattering photons from laser~1. This confirms, that after completion of
the second STIRAP pulse we only detect molecules that were previously in state $|g\rangle$. We
observe an efficiency of 75\% for the full cycle of conversion into state $|g\rangle$ and
back. Fig.~\ref{fig:stirapplot}c shows how this efficiency depends critically on the two
photon detuning $\delta$.

\begin{figure}
\includegraphics{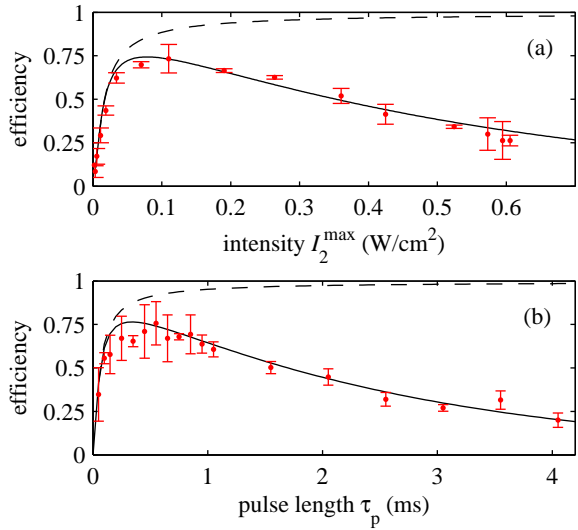}
\caption{ Efficiency for population transfer from state $|a\rangle$ to state $|g\rangle$ and
back with two STIRAP pulses. (a) Efficiency versus the laser intensity $I^\text{max}_{2}$
(fixed pulse length of $\tau_\text{p}=200\,$\textmu{}s). (b) Efficiency versus the pulse
length $\tau_\text{p}$ (fixed laser intensity $I^\text{max}_{2}= 44\,$mW/cm$^2$). For (a) and
(b) the intensity ratio $I^\text{max}_{2}/I^\text{max}_{1}=1/3.2$. The lines are from
calculations without free parameters using Eqs.~\ref{equ:formel}. Setting $\gamma_{a} =
\gamma_{g} =  0$, the efficiency would reach unity for a fully adiabatic transfer (dashed
lines). Using for $\gamma_{a}, \gamma_{g}$ the experimentally determined values, the
calculations (solid lines) are in good agreement with the data.}
  \label{fig:combined}
\end{figure}

In Fig.~\ref{fig:combined} we investigate further the complete STIRAP cycle efficiency as a
function of the laser intensity and pulse length. In these measurements we use pulses with the
same shape as in Fig.~\ref{fig:stirapplot}a, which are rescaled to adjust pulse time
$\tau_\text{p}$ and laser intensity. Again, for the best settings we reach an efficiency of
about $75\%$ for the two pulses, which corresponds to a transfer efficiency to state
$|g\rangle$ of about $87\%$. The dependence of the efficiency on intensity and pulse length
can be qualitatively understood as follows. For short pulse lengths or low intensities, the
dark state cannot adiabatically follow the STIRAP pulse, resulting in a low transfer
efficiency. For very long pulse lengths and high intensities the losses due to an imperfect
dark state become dominant, also resulting in a low transfer efficiency. Thus in order to find
an optimum value for the transfer efficiency there is a trade-off between adiabaticity and
inelastic photon scattering.

We are also able to quantitatively understand our data by using a three level model. It
describes the evolution of the quantum mechanical probability amplitudes $a$, $b$, and $g$ for
a molecule in the respective states $|a\rangle$, $|b\rangle$ and $|g\rangle$ in terms of the
following set of differential equations:
\begin{equation}
  \begin{aligned}
    i \dot{a} &= (- i\gamma_{a}/2) a -\tfrac{1}{2}\Omega_{1} b,\\
    i \dot{b} &= \left[(\Delta+\delta)-i\gamma_b/2 \right] b - \tfrac{1}{2}(\Omega_{1} a + \Omega_{2} g),\\
    i \dot{g} &= (\delta - i\gamma_{g}/2) g - \tfrac{1}{2}\Omega_{2} b.
    \label{equ:formel}
  \end{aligned}
\end{equation}
Here, the Rabi frequencies $\Omega_{1}, \Omega_{2}$, the detunings $\Delta$ and $\delta$ and
the decay rates $\gamma_{a}, \gamma_{b}, \gamma_{c}$ are defined as shown in
Fig.~\ref{fig:raman-scheme}. After experimentally determining $\Omega_{1}, \Omega_{2}$ and
$\gamma_{a},  \gamma_{g}$ and using $\gamma_b = 2\pi\times12\,\text{MHz}$, we are able to
consistently describe all data in Figs.~2, 3 and 4 with a single set of parameters. From
one-photon and two-photon scans (as e.g. in Fig.~\ref{fig:autler}) we obtain $\Omega_1 =
2\pi\times2.9\,\text{MHz} (I_1 / \text{(W\,cm}^{-2}))^{1/2} $ and $ \Omega_2
=2\pi\times6.0\,\text{MHz} (I_2 / \text{(W\,cm}^{-2}))^{1/2}$. The effective decay rates
$\gamma_{a},  \gamma_{g}$ are intensity dependent and are mainly due to the off-resonant
coupling of $|a\rangle$  with laser~2 and $|g\rangle$ with laser~1. We determine $\gamma_{a}$
($\gamma_{g}$) by shining laser~2 (laser~1) on the molecules in state $|a\rangle$
($|g\rangle$) and measuring the off-resonant losses. We find that $\gamma_{a}/I_2 = 2\pi
\times 0.72\,\text{kHz}/(\text{W\,cm}^{-2})$ and $\gamma_{g}/I_1 = 2\pi \times
0.40\,\text{kHz}/(\text{W\,cm}^{-2})$. From these independent measurements, all parameters of
Eqs.~\ref{equ:formel} are determined without further adjustable parameters. In the
calculations shown in Fig.~\ref{fig:stirapplot}b,c and Fig.~\ref{fig:combined} the time
dependent pulse shape (see Fig.~\ref{fig:stirapplot}a) are included. The agreement between
theory and experiment is very good.

\begin{figure}
\includegraphics[width=0.9 \columnwidth]{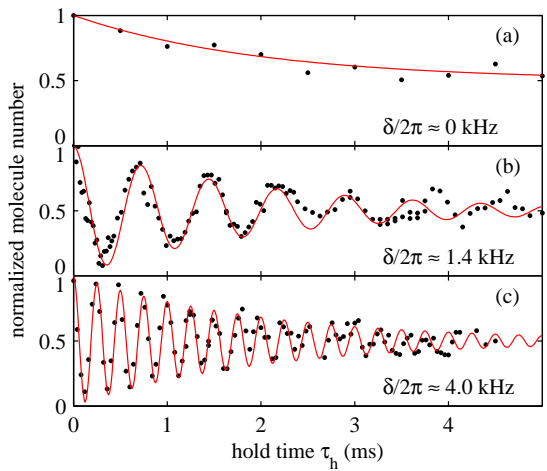}
\caption{Coherence of the $(|a\rangle - |g\rangle$) superposition state. Shown is the molecule
number in state $|a\rangle$ as a function of holding time $\tau_\text{h}$ for different
detunings $\delta$ as indicated. The oscillations indicate coherent flopping of the molecular
superposition state between the dark and a bright state. The lines are given by $0.5
\exp(-\tau_\text{h}/\tau) \cos(\delta \tau_\text{h}) + 0.5$, with a damping time $\tau =
2\,$ms.} \label{fig:coherenz}
\end{figure}

During STIRAP the molecules are in a quantum superposition, $\Omega_{2}|a\rangle -
\Omega_{1}|g\rangle$, which is dynamically evolving with the intensities of the lasers. In
order to probe the coherence of this superposition, we perform a Ramsey-type experiment. First
we create a dark superposition state with equal population in the two states, $|a\rangle -
|g\rangle$, by going halfway into the first STIRAP pulse of Fig. \ref{fig:stirapplot}a. We
then simultaneously switch off both STIRAP lasers for a variable holding time $\tau_\text{h}$,
after which we retrace in time the same STIRAP half pulse. As a result we observe oscillations
in the number of molecules in level $|a\rangle$ as a function of the holding time
$\tau_\text{h}$ (see Fig.~\ref{fig:coherenz}). During the hold time, the superposition state
freely evolves, coherently flopping between the dark and a bright state with a frequency equal
to the two-photon detuning $\delta$. At the end of the hold time, when we switch on again the
STIRAP lasers, the dark state is transferred back to state $|a\rangle$ whereas the bright
state will be immediately destroyed by the light and leads to complete loss of the
corresponding molecules. The observed oscillations are exponentially damped on a time scale of
about 2\,ms. This damping can be explained by a magnetic field inhomogeneity of about 20\,mG
over the molecular cloud, which leads to a spatial variation of $2\pi\times250\,$Hz in the
two-photon detuning $\delta$.

Furthermore, we have performed lifetime measurements of the molecules in state $|g\rangle$ by
varying the hold time $\tau_\text{h}$ between the two STIRAP pulses (see Fig. 2a). At a
lattice depth of 60 $E_r$ for molecules, we observe a long  lifetime of 0.8\,s (assuming
exponential decay), which is longer than the lifetime of 0.4\,s for Feshbach molecules in
state $|a\rangle$. At these deep lattices molecular decay is exclusively due to inelastic
scattering of lattice photons.

To conclude, using STIRAP  we have demonstrated a coherent transfer of a molecular quantum gas
from a weakly bound molecular level to a more deeply bound molecular level with a high
transfer efficiency of 87\%. The method can be extended in a straight forward manner to create
arbitrarily deeply bound molecules. With a \emph{single} STIRAP pulse all vibrational levels
down to level $X^1 \Sigma_g^+ \ (v = 116) / a^3 \Sigma_u^+ \ (v = 32)$ should be easily
reached since the Franck-Condon factors to state $|b\rangle$  are of similar order as for
level $|g\rangle$~\cite{Koc06}. This includes the level $X^1 \Sigma_g^+ \ (v = 119) / a^3
\Sigma_u^+ \ (v = 35)$ with its binding energy of 30\,GHz$\times h$, from which the
vibrational ground state $X^1 \Sigma_g^+ \ (v = 0)$ of the singlet potential can be reached
with two additional Raman (or STIRAP) transitions~\cite{Jak02}. Thus STIRAP is a promising
tool for the creation of a molecular BEC in the molecular ground state.

We thank Christiane Koch for providing Franck-Condon factors and Helmut Ritsch for valuable
discussions. This work was supported by the Austrian Science Fund (FWF) within SFB 15 (project
part 17) and the Cold Molecules TMR Network under Contract No.HPRN-CT-2002-00290. PvdS
acknowledges support within the ESF-program QUDEDIS during his stay in Innsbruck.

\input{bibliography}
\end{document}

%% file: figure1.tex
\begin{picture}(0,0)%
\includegraphics{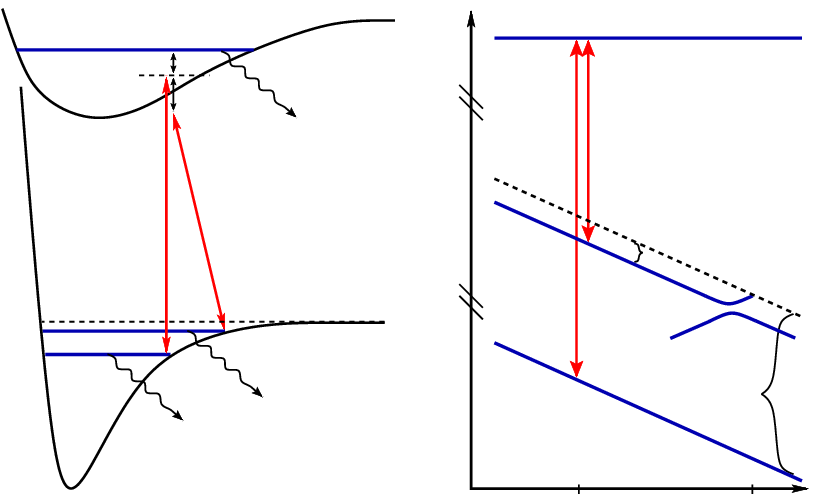}%
\end{picture}%
\setlength{\unitlength}{987sp}%
\begingroup\makeatletter\ifx\SetFigFont\undefined%
\gdef\SetFigFont#1#2#3#4#5{%
  \reset@font\fontsize{#1}{#2pt}%
  \fontfamily{#3}\fontseries{#4}\fontshape{#5}%
  \selectfont}%
\fi\endgroup%
\begin{picture}(15829,9970)(1082,-14487)
\put(10726,-4786){\makebox(0,0)[lb]{\smash{{\SetFigFont{8}{6.0}{\rmdefault}{\mddefault}{\updefault}{\color[rgb]{0,0,0}(b)}%
}}}}
\put(13518,-12167){\makebox(0,0)[lb]{\smash{{\SetFigFont{8}{6.0}{\rmdefault}{\mddefault}{\updefault}{\color[rgb]{0,0,0}637MHz}%
}}}}
\put(13901,-9454){\makebox(0,0)[lb]{\smash{{\SetFigFont{8}{6.0}{\rmdefault}{\mddefault}{\updefault}{\color[rgb]{0,0,0}24MHz}%
}}}}
\put(14581,-14461){\makebox(0,0)[lb]{\smash{{\SetFigFont{8}{6.0}{\rmdefault}{\mddefault}{\updefault}{\color[rgb]{0,0,0}1007.4G}%
}}}}
\put(9551,-6800){\rotatebox{90.0}{\makebox(0,0)[lb]{\smash{{\SetFigFont{8}{6.0}{\rmdefault}{\mddefault}{\updefault}{\color[rgb]{0,0,0}energy/$h$}%
}}}}}
\put(16726,-14090){\makebox(0,0)[lb]{\smash{{\SetFigFont{8}{6.0}{\rmdefault}{\mddefault}{\updefault}{\color[rgb]{0,0,0}B}%
}}}}
\put(11746,-14472){\makebox(0,0)[lb]{\smash{{\SetFigFont{8}{6.0}{\rmdefault}{\mddefault}{\updefault}{\color[rgb]{0,0,0}973G}%
}}}}
\put(1463,-4786){\makebox(0,0)[lb]{\smash{{\SetFigFont{8}{6.0}{\rmdefault}{\mddefault}{\updefault}{\color[rgb]{0,0,0}(a)}%
}}}}
\put(5100,-8025){\makebox(0,0)[lb]{\smash{{\SetFigFont{8}{6.0}{\familydefault}{\mddefault}{\updefault}{\color[rgb]{0,0,0}laser 1}%
}}}}
\put(3436,-9338){\makebox(0,0)[lb]{\smash{{\SetFigFont{8}{6.0}{\familydefault}{\mddefault}{\updefault}{\color[rgb]{0,0,0}$\Omega_{2}$}%
}}}}
\put(5150,-8738){\makebox(0,0)[lb]{\smash{{\SetFigFont{8}{6.0}{\familydefault}{\mddefault}{\updefault}{\color[rgb]{0,0,0}$\Omega_{1}$}%
}}}}
\put(2352,-5977){\makebox(0,0)[lb]{\smash{{\SetFigFont{8}{6.0}{\familydefault}{\mddefault}{\updefault}{\color[rgb]{0,0,0}$|b\rangle$}%
}}}}
\put(2382,-11847){\makebox(0,0)[lb]{\smash{{\SetFigFont{8}{6.0}{\familydefault}{\mddefault}{\updefault}{\color[rgb]{0,0,0}$|g\rangle$}%
}}}}
\put(2425,-10450){\makebox(0,0)[lb]{\smash{{\SetFigFont{8}{6.0}{\familydefault}{\mddefault}{\updefault}{\color[rgb]{0,0,0}$|a\rangle$}%
}}}}
\put(4501,-12976){\makebox(0,0)[lb]{\smash{{\SetFigFont{8}{6.0}{\familydefault}{\mddefault}{\updefault}{\color[rgb]{0,0,0}$ \gamma_g$}%
}}}}
\put(6150,-12500){\makebox(0,0)[lb]{\smash{{\SetFigFont{8}{6.0}{\familydefault}{\mddefault}{\updefault}{\color[rgb]{0,0,0}$ \gamma_a$}%
}}}}
\put(6800,-7122){\makebox(0,0)[lb]{\smash{{\SetFigFont{8}{6.0}{\familydefault}{\mddefault}{\updefault}{\color[rgb]{0,0,0}$\gamma_b$}%
}}}}
\put(4540,-6529){\makebox(0,0)[lb]{\smash{{\SetFigFont{8}{6.0}{\familydefault}{\mddefault}{\updefault}{\color[rgb]{0,0,0}$\delta$}%
}}}}
\put(4242,-5887){\makebox(0,0)[rb]{\smash{{\SetFigFont{8}{6.0}{\familydefault}{\mddefault}{\updefault}{\color[rgb]{0,0,0}$\Delta$}%
}}}}
\put(12956,-10157){\makebox(0,0)[lb]{\smash{{\SetFigFont{8}{6.0}{\familydefault}{\mddefault}{\updefault}{\color[rgb]{0,0,0}$|a\rangle$}%
}}}}
\put(12972,-5809){\makebox(0,0)[lb]{\smash{{\SetFigFont{8}{6.0}{\familydefault}{\mddefault}{\updefault}{\color[rgb]{0,0,0}$|b\rangle$}%
}}}}
\put(12575,-7838){\makebox(0,0)[lb]{\smash{{\SetFigFont{8}{6.0}{\familydefault}{\mddefault}{\updefault}{\color[rgb]{0,0,0}$\Omega_{1}$}%
}}}}
\put(12778,-12792){\makebox(0,0)[lb]{\smash{{\SetFigFont{8}{6.0}{\familydefault}{\mddefault}{\updefault}{\color[rgb]{0,0,0}$|g\rangle$}%
}}}}
\put(11311,-10463){\makebox(0,0)[lb]{\smash{{\SetFigFont{8}{6.0}{\familydefault}{\mddefault}{\updefault}{\color[rgb]{0,0,0}$\Omega_{2}$}%
}}}}
\put(2450,-8678){\makebox(0,0)[lb]{\smash{{\SetFigFont{8}{6.0}{\familydefault}{\mddefault}{\updefault}{\color[rgb]{0,0,0}laser 2}%
}}}}
\end{picture}%

%% file: stirap.bbl
\begin{thebibliography}{99}

\bibitem{review-nature} For overviews, see {\em Ultracold matter}, Nature Insight, Nature (London) {\bf 416}, 205--246 (2002).

\bibitem{Chi05} C. Chin {\em et al.},
%T. Kraemer, M. Mark, J. Herbig, P. Waldburger, H.-C. Nägerl, and R. Grimm,
Phys. Rev. Lett. {\bf 94}, 123201 (2005).

\bibitem{Sta06} P. Staanum, S.D. Kraft, J. Lange, R. Wester, and M. Weidemüller,
% Experimental Investigation of Ultracold Atom-Molecule Collisions
Phys. Rev. Lett. {\bf 96}, 023201 (2006).

\bibitem{Zah06} N. Zahzam, T. Vogt, M. Mudrich, D. Comparat, and P. Pillet,
% Atom-Molecule Collisions in an Optically Trapped Gas
Phys. Rev. Lett. {\bf 96}, 023202 (2006).

% Quantum computation
\bibitem{Mil02} D. deMille, Phys. Rev. Lett. {\bf 88}, 067901 (2002).

\bibitem{Doy04} For an overview, see  J. Doyle, B. Friedrich, R.V. Krems, and F. Masnou-Seeuws,
% Quo vadis, cold molecules?
Eur. Phys. J. D {\bf 31}, 149 (2004).

\bibitem{Jon06} For an overview, see  K.M. Jones, E. Tiesinga, P.D. Lett, and P.S. Julienne,
% Ultracold photoassociation spectroscopy: Long-range molecules and atomic scattering
Rev. Mod. Phys. {\bf 78}, 483 (2006).

\bibitem{Koh06} For an overview, see T. K\"ohler and K. Goral, and P.S. Julienne,
% Production of cold molecules via magnetically tunable Feshbach resonances
% submitted to Rev. Mod. Phys.
cond-mat/0601420.

\bibitem{Jak02}
% Creation of a Molecular Condensate by Dynamically Melting a Mott  Insulator
D. Jaksch, V. Venturi, J. I. Cirac, C. J. Williams, and P. Zoller, Phys. Rev. Lett. {\bf 89}, 040402 (2002).

\bibitem{Sag05}
% Optical Production of Ultracold Polar Molecules
J.M. Sage, S. Sainis, T. Bergeman, and D. DeMille, Phys. Rev. Lett. {\bf 94}, 203001 (2005).

\bibitem{Ber98} K. Bergmann, H. Theuer, and B.W. Shore, Rev. Mod. Phys. {\bf 70}, 1003 (1998).

\bibitem{Wynar} R.~Wynar, R.S.~Freeland, D.J.~Han, C.~Ryu, and D.J.~Heinzen,
% Moleculas in a Bose Einstein condensate
Science {\bf 287}, 1016 (2000).

\bibitem{Rom}   T. Rom {\em et al.},
%  T. Rom, T. Best, O. Mandel, A. Widera, M. Greiner,
% T.W. H\¨ansch, and I. Bloch,
% State selective production of molecules in optical lattices
Phys. Rev. Lett. {\bf 93}, 073002 (2004).

\bibitem{Win05} K. Winkler {\em et al.},
% G. Thalhammer, M. Theis, H. Ritsch, R. Grimm, and J. Hecker Denschlag,
Phys. Rev. Lett. {\bf 95}, 063202 (2005).

\bibitem{Tha05} G. Thalhammer {\em et al.}, Phys. Rev. A {\bf73}, 033403 (2005).

\bibitem{Tha06}
% Long-lived Feshbach molecules in a 3D optical lattice
G. Thalhammer {\em et al.},
%, K. Winkler, F. Lang, S. Schmid, R. Grimm, and J. Hecker Denschlag
Phys. Rev. Lett. {\bf 96}, 050402 (2006).

\bibitem{Vol03} T. Volz, S. Dürr, S. Ernst, A. Marte, and G. Rempe,
%Characterization of elastic scattering near a Feshbach resonance in 87Rb
Phys. Rev. A {\bf 68}, 010702(R) (2003).

\bibitem{Fioretti}  A. Fioretti {\em et al.},
% C. Amiot, C.M. Dion, O. Dulieu, M. Mazzoni, G. Smirne, and C. Gabbanini,
Eur. Phys. J. D {\bf 15}, 189 (2001).

\bibitem{Koc06} C. Koch, private communication.



 \end{thebibliography}
